\documentclass[prb,12pt]{revtex4}

\usepackage{graphicx}
\usepackage{amsmath}
\usepackage{amssymb}
\begin{document}
\title{Effect of ionization/recombination processes on the electrical interactions between positively charged particles in highly collisional plasmas}

\author{ M. Chaudhuri, S. A. Khrapak, and G. E. Morfill}

\affiliation{Max-Planck Institut f\"ur extraterrestrische Physik,
D-85741 Garching, Germany}

\date{\today}

\begin{abstract}
The effect of ionization and recombination processes on the electrical interactions between a pair of small charged particles in highly collisional plasmas is discussed. In particular, it is shown that these processes suppress the long-range attraction between positively charged particles. The condition corresponding to the vanishing of attraction is derived. The role of the effect for conditions of existing experiments is estimated.
\end{abstract}

\maketitle


The shape of the interparticle interaction potential is a very important factor which determines various physical phenomena in complex (dusty) plasmas (as well as in other interacting particle systems). Different mechanisms of attraction and repulsion between the charged particles immersed in plasmas have been discussed in the literature.\cite{ShuklaIOP,Morfill_Springer,Fortov2005PR,Fortov2004UFN,KhrapakCPP} At short distances electrical repulsion between like-charged particles dominates.
In isotropic conditions this repulsion can be approximated by the Debye-H\"uckel (Yukawa) potential $U(r) = (Q^2/r)\exp(-r/\lambda)$, where $Q$ is the particle charge and $\lambda$ is the plasma screening length. At longer distances various effects can contribute to the interparticle interaction. Among these is the so-called ``ion shadowing'' interaction,~\cite{IgnatovPPR1996,TsytovichCPPCF1996} which can generate an attractive branch of the interaction potential.~\cite{CP,Note} A similar effect known as ``neutral shadowing'' can lead to additional attractive or repulsive contribution to the interparticle interaction potential, depending on the sign of the temperature difference between the particle surface and surrounding gas.~\cite{TsytovichNS,CEC} Even the potential of electrical interaction between the charged particles can have complicated structures at large interparticle distances. In general it is not screened exponentially, but has an inverse power-law long-range asymptote as long as plasma production and loss processes in the particles vicinity are neglected.~\cite{KhrapakCPP,KhrapakPRL2008} The potential decays as $\propto 1/r^2$ in collisionless plasmas~\cite{Alpert,TsytovichUFN} and as $\propto 1/r$ in highly collisional plasmas.\cite{KhrapakPRL2008,Su,KhrapakPoP2006} Moreover, electrical interaction can exhibit long-range repulsion or attraction, depending on the sign of the particle charge.

As long as collective effects relevant to dense dust clouds are neglected, two negatively charged particles repel each other electrically at any distance. However, two positively charged particles can attract each other at long distances. The possibility of attraction between individual positively charged emitting particles was pointed out by Delzanno {\it et al}. who performed a simulation study of the particle charging in a collisionless plasma, taking into account the effect of thermionic emission from the particle surface.~\cite{DelzanoPRL-92-035002-2004} They constructed an analytical theory to describe this effect in collisionless plasmas.~\cite{DelzanoPoP-12-062102-2005} Later on, an analytical approach to calculate the attractive part of the potential between positively charged emitting particles was developed for the highly collisional plasma regime.~\cite{KhrapakPRL-99-055003-2007} The physical mechanism of this attraction was discussed in Ref.~\onlinecite{KhrapakPRL-99-055003-2007} and it was suggested there that attraction can explain the formation of ordered structures in the ``dusty combustion experiment''.~\cite{FortovPRE1996} However, an important effect -- ionization and recombination in the background plasma -- was neglected in this theoretical study.

As has been recently demonstrated, the potential of electrical interaction between the particles is sensitive to the strength and exact nature of the plasma production and loss processes.~\cite{Filippov2007a,Chaudhuri2008a} So far, only the case of negatively charged particles has been considered in literature. The purpose of this Brief Communication is to estimate the effect of ionization and recombination processes on the attractive branch of the interaction potential between a pair of small positively charged particles in plasmas. In particular, we find that ionization suppresses the attractive part of the potential and, when the ionization rate becomes sufficiently high, the attractive branch disappears. We identify the condition when this happens. We also estimate the importance of ionization/recombination processes for the conditions of  the ``dusty combustion experiment'', and demonstrate that here they play only a negligible role.

We assume, below, that electron impact ionization is responsible for the plasma production, while plasma losses are due to the electron-ion volume recombination.~\cite{Filippov2007a,Chaudhuri2008a}
In the limit of highly collisional plasma ($\ell_{i(e)} \ll \lambda_{\rm D}$), both the ion and electron components are mobility controlled and can be described by hydrodynamic equations. Here $\ell_{i(e)}$ is the mean free path of ions (electrons), $\lambda_{\rm D} = (\lambda_{{\rm D}e}^{-2} + \lambda_{{\rm D}i}^{-2})^{-1/2}$ is the linearized Debye radius, $\lambda_{{\rm D}i(e)} = \sqrt{T_{i(e)}/4\pi n_0e^2}$ is the ion (electron) Debye radius, $T_{i(e)}$ is the ion (electron) temperature, and  $n_i\simeq n_e\simeq n_0$ is the unperturbed plasma density. Due to the presence of fast thermalization processes in highly collisional plasma, we can assume that the electron temperature (of both emitted and background electrons) is uniform. Finally, we consider the potential distribution around an individual particle and the resulting interaction between a pair of particles, thus neglecting any collective effects including plasma losses on the particle component.

The distribution of electrical potential around an individual particle can be calculated combining the continuity and momentum equations for electrons and ions in the hydrodynamic approximation with the Poisson equation. The result is\cite{Filippov2007a,Chaudhuri2008a}
\begin{equation} \label{isotropicpotential}
\phi(r) = (Q_+ /r)\exp(-rk_+) + (Q_-/r)\exp(-rk_-),
\end{equation}
where
\begin{equation} \label{charge}
Q_{\pm} = \mp\frac{Q\left[k_{\mp}^2 - k_{\rm D}^2
- \left(eJ_0/QD_i\right)\left(1 - D_i/D_e\right)\right]}{k_+^2 - k_-^2}
\end{equation}
and
\begin{equation}\label{screeninglength}
2k_\pm^2 = k_{\rm D}^2 +\nu_I/D_i \pm \sqrt{\left(k_{\rm D}^2 +
\nu_I/D_i\right)^2 - 4\nu_I\left(k_{{\rm D}e}^2/D_i + k_{{\rm D}i}^2/D_e\right)}.
\end{equation}
Here $D_{i(e)}=v_{Ti(e)}\ell_{i(e)}$, $v_{Ti(e)}=\sqrt{T_{i(e)}/m_{i(e)}}$, $m_{i(e)}$, and $k_{{\rm D}i(e)} = \lambda_{{\rm D}i(e)}^{-1}$ are the diffusion coefficient, thermal velocity, mass, and inverse Debye radius for ions (electrons) respectively. The inverse linearized Debye radius is $k_{\rm D} = \sqrt{k_{{\rm D}e}^2 + k_{{\rm D}i}^2}$. In the stationary state the electron and ion fluxes collected by the particle are equal to each other, $J_i=J_e=J_0$. The total electron flux $J_e$ includes both the contribution from the electrons absorbed on the particle surface and those emitted from the surface. We assume here that electron emission is quite significant, so that the particle charge is positive, $Q>0$. The ionization rate $\nu_I$ is related to the recombination constant $\beta$ via $\nu_I=\beta n_0$.

Equation (\ref{isotropicpotential}) demonstrates that in the considered case the potential is screened exponentially, but unlike in the Debye-H\"{u}ckel theory the potential contains the superposition of the two exponentials with different inverse screening lengths (screening parameters) $k_+$ and $k_-$. Both of these screening parameters depend on the strength of plasma production through the ionization frequency $\nu_I$. The effective charges $Q_+$ and $Q_-$ both depend on the strength of plasma production and plasma fluxes collected by the particle. The short range part of the potential is determined by the first term in Eq. (\ref{isotropicpotential}) with the effective charge $Q_+$ and the larger screening parameter $k_+$. It can be easily shown that $Q_+>0$ and therefore this part is always repulsive. On the other hand, the long range asymptote of the potential is determined by the smaller screening parameter $k_-$ with the effective charge $Q_-$. The long-range attraction operates when $Q_-<0$. From Eq.~(\ref{charge}) we immediately get an approximate condition for the existence of attraction. It can be written as
\begin{equation}
J_0\gtrsim \nu_I(Q/e).
\end{equation}
It is obvious from the obtained inequality that attraction between two positively charged particles exists when the ionization strength is low enough. Otherwise, the potential of electrical interaction is purely repulsive.

Note that if we neglect completely the ionization and recombination processes ($\nu_I=0, \beta=0$) the potential (\ref{isotropicpotential}) reduces to that obtained previously [Eq. (8) of Ref.~\onlinecite{KhrapakPRL-99-055003-2007}]. The effect of ionization is small when $k_{\rm D}^2\gtrsim \nu_I/D_i$. In this case the smaller screening parameter which determines the global screening is $k_-^2\simeq k_{{\rm D}e}^2(\nu_I/k_{\rm D}^2D_i)[1+(k_{{\rm D}i}^2D_i/k_{{\rm D}e}^2D_e)]$. Usually $(k_{{\rm D}i}^2D_i/k_{{\rm D}e}^2D_e)\ll 1$ and, therefore, $k_-\simeq k_{{\rm D}e}\sqrt{\nu_I/k_{\rm D}^2D_i}$.
This yields a lengths scale $R\sim \lambda_{{\rm D}e}\sqrt{k_{\rm D}^2D_i/\nu_I}\gtrsim \lambda_{{\rm D}e}$ above which ionization and recombination in plasma should be taken into account. For $r\lesssim R$ the model neglecting ionization and recombination processes is applicable.


As suggested in Ref.~\onlinecite{KhrapakPRL-99-055003-2007}, electrical attraction between positively charged emitting particles can explain the formation of ordered structures in the ``dusty combustion experiment'' by Fortov {\it et al}.~\cite{FortovPRE1996,FortovJETPLett1996,KhodataevPRE1998} In this experiment the dust particles (CeO$_2$ grains of radius $a \simeq 0.4$ $\mu$m) were injected into a laminar air spray at atmospheric pressure and temperature $T \sim 1700-2200$ K created by a two-flame Meeker burner. The complex (dusty) plasma constituents were air, electrons, Na$^+$ ions and CeO$_2$ particles, all in thermal equilibrium. The grains were charged positively by emitting thermal electrons up to $Q\sim 10^2e$. At $T \sim 1700$ K the particle component formed a short range ordered structure, with the pronounced first maximum in the pair correlation function. This phenomena can be attributed to the presence of the weak attractive part of the interparticle interaction potential, with the minimum of the potential corresponding approximately to the position of the first maximum in the pair correlation function.~\cite{KhrapakPRL-99-055003-2007}  However, the effects of ionization and recombination were not taken into account in Ref.~\onlinecite{KhrapakPRL-99-055003-2007}. Let us therefore estimate their importance for the considered experimental conditions.

Let us first estimate the important plasma parameters. Some of them were already estimated in Ref.~\onlinecite{KhrapakPRL-99-055003-2007}. According to these estimates for $T \sim 1700$ K and $n_e \sim 7 \times 10^{10}$ cm$^{-3}$ we have  $\ell_i\sim 0.06$ $\mu$m, $\lambda_{\rm D}\sim 10$ $\mu$m, and $v_{Ti}\sim 8\times 10^4$ cm/s. This yields $D_i\sim 0.5$ cm$^2$/s for the ion diffusion coefficient. The next step is to estimate the ionization rate $\nu_I$ for the condition of the experiment. We use the Saha equation,~\cite{Saha} $(n_en_i/N)=2(g_i/g_a)(2\pi m_e T/h^2)^{3/2}\exp(-I/T)$, to determine the density of Na atoms. Using $g_i=1$, $g_a=2$, and $I\simeq 5.1$ eV  (here $I$ denotes the ionization potential of Na atoms) we get $N\sim 2\times 10^{16}$ cm$^{-3}$. Then, an approximate expression for the electron impact ionization frequency,~\cite{raizer} $\nu_I\simeq \sqrt{8/\pi}Nv_{Te}C_I(I+2T)\exp(-I/T)$, yields $\nu_I\sim 4\times 10^{-7}$ s$^{-1}$, where $C_I\simeq 10^{-16}$ cm$^2$/eV is used.~\cite{Johnston} Finally, for the ion flux collected by the particle we get~\cite{KhrapakPRL-99-055003-2007} $J_0\simeq 4\pi a n_0D_iz \exp(-z)\sim 4\times 10^6$ s$^{-1}$, where $z=Qe/aT\sim 2.5$ is the dimensionless particle surface potential. Looking at these numbers we immediately see that $k_{\rm D}^2\gg \nu_I/D_i$ and $J_0\gg \nu_I(Q/e)$. This implies that for this particular experiment, the effect of ionization and recombination plays a negligible role and the conclusions of Ref.~\onlinecite{KhrapakPRL-99-055003-2007} remain unaffected. We would like to point out that, in the experiment discussed above, plasma losses on the particle component can play a considerable role due to the large plasma fluxes directed to the particles and non-vanishing particle concentration. This effect along with potentially important collective effects is, however, beyond the scope of this Brief Communication and will be considered separately.

To summarize, ionization and recombination processes constitute an important factor which affects the distribution of electrical potential around an individual particle in plasmas and therefore is important for electrical interaction between the particles. We have shown here that ionization/recombination processes can suppress the attractive branch of the interaction potential between a pair of positively charged particles, and derived the condition when this attraction vanishes. According to our estimates, ionization/recombination effects are negligible for the conditions of the ``dusty combustion experiment''. However, we cannot exclude that in other regimes they can play a dominant role.

This work was partly supported by DLR under Grant 50WP0203 (Gef\"{o}rdert von der Raumfahrt-Agentur des Deutschen Zentrums f\"{u}r Luft und Raumfahrt e. V. mit Mitteln des Bundesministeriums f\"{u}r Wirtschaft und Technologie aufgrund eines Beschlusses des Deutschen Bundestages unter dem F\"{o}rderkennzeichen 50 WP 0203.).

\end{document}